# Simulation vs. Equivalence

Zoltán Ésik[1,*] and Andreas Maletti[2,†]

[1]Dept. of Computer Science, University of Szeged
Árpád tér 2, 6720 Szeged, Hungary
ze@inf.u-szeged.hu

[2]Dept. de Filologies Romàniques, Universitat Rovira i Virgili
Avinguda de Catalunya 35, 43002 Tarragona, Spain
andreas.maletti@urv.cat

**Abstract**

For several semirings $S$, two weighted finite automata with multiplicities in $S$ are equivalent if and only if they can be connected by a chain of simulations. Such a semiring $S$ is called "proper". It is known that the Boolean semiring, the semiring of natural numbers, the ring of integers, all finite commutative positively ordered semirings and all fields are proper. The semiring $S$ is Noetherian if every subsemimodule of a finitely generated $S$-semimodule is finitely generated. First, it is shown that all Noetherian semirings and thus all commutative rings and all finite semirings are proper. Second, the tropical semiring is shown not to be proper. So far there has not been any example of a semiring that is not proper.

## 1 Introduction

In this paper, we consider weighted (finite) automata [2, 5] with multiplicities (or weights) in a semiring $S$. A weighted automaton over a finite alphabet $\Sigma$ with multiplicities in $S$ defines a *rational series* [2, 5, 9] in the semiring $S\langle\!\langle \Sigma^* \rangle\!\rangle$ of all formal series over $\Sigma$ with coefficients in $S$. Two automata are termed *equivalent* if they define the same rational series.

In [3, 4], a notion of morphism between automata was introduced in order to relate equivalent automata. These morphisms, called "simulations" preserve the equivalence of automata. It has been demonstrated that for many semirings, any two equivalent automata over any finite alphabet can be connected by a finite chain of simulations. Semirings with this property for all finite alphabets include the Boolean semiring [3], any finite commutative positively ordered semiring [4], any field [1], the semiring $\mathbb{N}$ of natural numbers and the ring of integers [1]. Such semirings are called *proper*. Until now, there has not been any example of a semiring that is not proper.

[*]Partially supported by grant no. K 75249 from the National Foundation of Hungary for Scientific Research and by the TÁMOP-4.2.2/08/1/2008-0008 program of the Hungarian National Development Agency.
[†]Supported by the Ministerio de Educación y Ciencia (MEC) grant JDCI-2007-760.

In this note, our aim is two-fold. First, we point out additional classes of proper semirings. We call a semiring $S$ Noetherian if every subsemimodule of a finitely generated $S$-semimodule is finitely generated. In Theorem 4.2 we show that any Noetherian semiring and thus any commutative ring and any finite semiring is proper. Then in Theorem 5.4 we prove that the tropical semiring [6, 7] used in many combinatorial optimization problems is *not* proper.

## 2 Semirings and semimodules

We recall from [6, 7] that a *semiring* $S = (S, +, \cdot, 0, 1)$ consists of a commutative monoid $(S, +, 0)$ and a monoid $(S, \cdot, 1)$ such that multiplication (or product) $\cdot$ distributes over addition (or sum) $+$, and moreover, $s \cdot 0 = 0 = 0 \cdot s$ for all $s \in S$. A semiring $S$ is called *commutative* if $ss' = s's$ for all $s, s' \in S$. (When writing expressions, we will follow the standard convention that multiplication has higher precedence than addition.) Examples of semirings include all fields and rings, all bounded distributive lattices including the 2-element lattice $\mathbb{B} = \{0, 1\}$, called the Boolean semiring, the semiring $\mathbb{N}$ of natural numbers, and the tropical semiring defined in Section 5. In order to avoid trivial situations, we will only consider nontrivial semirings in which $0 \neq 1$. When $S$ is a semiring, so is the collection $S^{n \times n}$ of all $n \times n$ matrices over $S$ with the usual operations and constants. We will identify any matrix in $S^{1 \times n}$ with the corresponding row vector, and any matrix in $S^{n \times 1}$ with the corresponding column vector.

If $S$ is a semiring, an $S$-*semimodule* is a commutative monoid $V = (V, +, 0)$ that is equipped with a (left) $S$-*action* $S \times V \to V$ with $(s, v) \mapsto sv$ subject to the usual laws:

$$(s + s')v = sv + s'v$$
$$s(v + v') = sv + sv'$$
$$(ss')v = s(s'v)$$
$$1v = v$$
$$s0 = 0$$
$$0v = 0$$

for all $s, s' \in S$ and $v, v' \in V$. Note that for any $m, n \geq 1$, the set $S^{m \times n}$ of $m \times n$ matrices equipped with the pointwise sum operation is an $S$-semimodule.

Suppose that $S$ is a semiring and $\Sigma$ is a finite alphabet. Let $\Sigma^*$ denote the free monoid of all words over $\Sigma$ including the empty word $\epsilon$. Recall from [2, 9] that a *formal series* over $\Sigma$ with multiplicities in $S$ is a function $s \colon \Sigma^* \to S$ written as a formal sum $\sum_{w \in \Sigma^*}(s, w)w$, where $(s, w) = s(w)$ for each $w \in \Sigma^*$.

The *support* supp($s$) of a series $s$ is $\{w \mid (s, w) \neq 0\}$. We let $S\langle\!\langle \Sigma^* \rangle\!\rangle$ denote the collection of all such series. Each $s \in S$ may be identified with the series mapping $\epsilon$ to $s$ and all nonempty words to 0. This defines the series 0 and 1. Also, each letter $a \in \Sigma$ may be identified with the series mapping $a$ to 1 and all other words of $\Sigma^*$ to 0. The sum and product operations are defined by

$$(s + s', w) = (s, w) + (s', w)$$
$$(ss', w) = \sum_{uv=w} (s, u)(s', v)$$

for all $s, s' \in S\langle\!\langle \Sigma^* \rangle\!\rangle$. It is well-known (see e.g. [2, 9]) that, equipped with the above operations and constants, $S\langle\!\langle \Sigma^* \rangle\!\rangle$ is a semiring. In particular, $\mathbb{B}\langle\!\langle \Sigma^* \rangle\!\rangle$ is isomorphic to the semiring $P(\Sigma^*)$ of all languages over $\Sigma$ equipped with set union as sum and concatenation as product. The canonical isomorphism of type $\mathbb{B}\langle\!\langle \Sigma^* \rangle\!\rangle \to P(\Sigma^*)$ maps a series in $\mathbb{B}\langle\!\langle \Sigma^* \rangle\!\rangle$ to its support.

Below we will denote by $S\langle \Sigma \rangle$ the set of all series in $S\langle\!\langle \Sigma^* \rangle\!\rangle$ whose support is a subset of $\Sigma$. Notice that each element of $S\langle \Sigma \rangle$ may be written as a linear combination $s_1 a_1 + \cdots + s_n a_n$, where $a_1, \ldots, a_n$ are the letters of $\Sigma$ and each coefficient $s_i$ is an element of $S$.

For later use we note that when $s_i$ with $i \in I$ are series in $S\langle\!\langle \Sigma^* \rangle\!\rangle$ with pairwise disjoint support, then it is possible to define their sum $\sum_{i \in I} s_i$ as the series $s$ with $(s, w) = (s_i, w)$ if $w \in \text{supp}(s_i)$ for some $i$ and $(s, w) = 0$ otherwise.

## 3 Automata and simulations

Suppose that $S$ is a semiring and $\Sigma$ is a finite alphabet. A (finite) *automaton* in $S\langle\!\langle \Sigma^* \rangle\!\rangle$ of dimension $n \geq 1$ is a triplet $\mathcal{A} = (\alpha, M, \beta)$, where $\alpha \in S^{1 \times n}$ is the *initial vector*, $M \in S\langle \Sigma \rangle^{n \times n}$ is the *transition matrix*, and $\beta \in S^{n \times 1}$ is the *final vector* of $\mathcal{A}$. Note that we may write $M$ as a finite sum $M = \sum_{a \in \Sigma} M_a a$ with $M_a \in S^{n \times n}$ for all $a \in \Sigma$, where $(M_a)_{ij}$ is the coefficient of $a$ in $M_{ij}$ for each $1 \leq i, j \leq n$.

Since each entry of $M$ is a series in $S\langle \Sigma \rangle$, for each $k \geq 0$ it holds that each entry of $M^k$ is a series whose support is included in $\Sigma^k = \{w \in \Sigma^* \mid |w| = k\}$, which is the set of all words in $\Sigma^*$ of length $k$. Thus, it is possible to define $M^*$ as the matrix

$$M^* = E_n + M + M^2 + \cdots = \sum_{i \geq 0} M^i$$

where $M^0 = E_n$ is the $n \times n$ identity matrix. It is easy to see that for any $1 \leq i, j \leq n$ and $w \in \Sigma^*$, we have $((M^*)_{ij}, w) = (M_w)_{ij}$, where $M_\epsilon = E_n$ and $M_{va} = M_v M_a$ for all $v \in \Sigma^*$ and $a \in \Sigma$. The *behavior* of the automaton $\mathcal{A}$ is defined as the series

$$|\mathcal{A}| = \alpha M^* \beta \in S\langle\!\langle \Sigma^* \rangle\!\rangle.$$

Alternatively, $|\mathcal{A}|$ is the series with $(|\mathcal{A}|, w) = \alpha M_w \beta$ for all $w \in \Sigma^*$. We say that automata $\mathcal{A}$ and $\mathcal{B}$ are *equivalent* if $|\mathcal{A}| = |\mathcal{B}|$. Note that an automaton in $\mathbb{B}\langle\!\langle \Sigma^* \rangle\!\rangle$ is just an ordinary finite nondeterministic automaton.

Simulations of automata were defined in [3, 4] in order to provide a structural characterization of equivalent automata.

**Definition 3.1.** *Let $\mathcal{A} = (\alpha, M, \beta)$ and $\mathcal{B} = (\gamma, N, \delta)$ be automata in $S\langle\!\langle \Sigma^* \rangle\!\rangle$ of dimension $m$ and $n$, respectively. We say that $X \in S^{m \times n}$ is a simulation $\mathcal{A} \to \mathcal{B}$, in notation $\mathcal{A} \to^X \mathcal{B}$, if*

$$\alpha X = \gamma$$
$$MX = XN$$
$$\beta = X\delta.$$

Note that the condition $MX = XN$ can be expressed in an equivalent way as $M_a X = X N_a$ for all $a \in \Sigma$.

If $X$ is a simulation $\mathcal{A} \to \mathcal{B}$, then $M^k X = X N^k$ and $M_w X = X N_w$ for all $k \geq 0$ and $w \in \Sigma^*$ and thus

$$\alpha M^k \beta = \alpha M^k X \delta = \alpha X N^k \delta = \gamma N^k \delta$$

for all $k \geq 0$, which proves that $\mathcal{A}$ and $\mathcal{B}$ are equivalent.

**Definition 3.2.** *Call a semiring $S$* proper *if for any finite alphabet $\Sigma$ and for any two automata $\mathcal{A}$ and $\mathcal{B}$ in $S\langle\!\langle \Sigma^* \rangle\!\rangle$ the following two conditions are equivalent:*

- *$\mathcal{A}$ and $\mathcal{B}$ are equivalent.*
- *There exists a finite chain of simulations connecting $\mathcal{A}$ and $\mathcal{B}$; i.e., there is a sequence of automata $\mathcal{C}_0, \ldots, \mathcal{C}_k$ with $k \geq 0$ such that $\mathcal{C}_0 = \mathcal{A}$, $\mathcal{C}_k = \mathcal{B}$, and for each $1 \leq i \leq k-1$, either there is a simulation $\mathcal{C}_i \to^{X_i} \mathcal{C}_{i+1}$ or a simulation $\mathcal{C}_{i+1} \to^{X_i} \mathcal{C}_i$.*

For later use we note:

**Lemma 3.3.** *If every finitely generated subsemiring of a semiring $S$ is contained in a proper subsemiring of $S$, then $S$ is proper.*

*Proof.* Suppose that $\mathcal{A} = (\alpha, M, \beta)$ and $\mathcal{B} = (\gamma, N, \delta)$ are equivalent automata in $S\langle\!\langle \Sigma^* \rangle\!\rangle$. Let $S_0$ denote the subsemiring of $S$ that is generated by the entries of $\alpha, \beta, \gamma, \delta$ and $M, N$. By assumption, $S_0$ is contained in a proper subsemiring $S_1$. Since $S_1$ is proper and $\mathcal{A}$ and $\mathcal{B}$ are equivalent automata in $S_1\langle\!\langle \Sigma^* \rangle\!\rangle$, they can be connected by a finite chain of simulations involving automata in $S_1\langle\!\langle \Sigma^* \rangle\!\rangle$ and simulation matrices over $S_1$. Trivially, those are automata in $S\langle\!\langle \Sigma^* \rangle\!\rangle$ and simulation matrices over $S$. □

In [1, 3, 4] several proper semirings $S$ have been identified, including the Boolean semiring [3], any finite positively ordered commutative semiring [4], any field [1], the semiring $\mathbb{N}$ of natural numbers and the ring $\mathbb{Z}$ of integers [1]. In Section 4, we will show that every Noetherian semiring and thus every commutative ring and every finite semiring is proper. Then, in Section 5, we will show that the tropical semiring is not proper.

# 4 Noetherian semirings

**Definition 4.1.** *We call a semiring $S$ Noetherian if for every finitely generated $S$-semimodule $A$, every subsemimodule of $A$ is finitely generated.*

**Theorem 4.2.** *Every Noetherian semiring is proper.*

*Proof.* Suppose that $S$ is a Noetherian semiring and $\Sigma$ is a finite alphabet. Let $\mathcal{A} = (\alpha, M, \beta)$ and $\mathcal{B} = (\gamma, N, \delta)$ be equivalent automata in $S\langle\!\langle \Sigma^* \rangle\!\rangle$ of dimension $m$ and $n$, respectively. Our aim is to construct an automaton $\mathcal{C} = (\kappa, R, \lambda)$ of dimension $p$, say, together with simulations $\mathcal{C} \to^X \mathcal{A}$ and $\mathcal{C} \to^Y \mathcal{B}$, where $X \in S^{p \times m}$ and $Y \in S^{p \times n}$.

To this end, for each $k \geq 0$, let

$$V_k = \{(\alpha_w, \gamma_w) \mid |w| \leq k\},$$

where $\alpha_w = \alpha M_w$ and $\gamma_w = \gamma N_w$. Moreover, let $\langle V_k \rangle$ be the subsemimodule of the $S$-semimodule of $(m+n)$-dimensional row vectors over $S$ generated by $V_k$. Since

$$\alpha_w \beta = \alpha M_w \beta = (|\mathcal{A}|, w) = (|\mathcal{B}|, w) = \gamma N_w \delta = \gamma_w \delta,$$

for all $w \in \Sigma^*$, we have $\alpha' \beta = \gamma' \delta$ for all $(\alpha', \gamma') \in \langle V_k \rangle$, $k \geq 0$ such that $\alpha' \in S^m$ and $\gamma' \in S^n$.

Since $V_k \subseteq V_{k+1}$, also $\langle V_k \rangle \subseteq \langle V_{k+1} \rangle$ for each $k \geq 0$. Since $S$ is Noetherian, it follows that there is some $k_0$ with $\langle V_{k_0} \rangle = \langle V_{k_0+1} \rangle$. Moreover, $\langle V_{k_0} \rangle$ is finitely generated. Let $\langle V_{k_0} \rangle = \langle \{(\alpha_1, \gamma_1), \ldots, (\alpha_p, \gamma_p)\} \rangle$, say, where $\alpha_i \in S^m$ and $\gamma_i \in S^n$ for all $i$. Since $V_{k_0+1} \subseteq \langle V_{k_0} \rangle$ for each letter $a \in \Sigma$, there is a matrix $R_a \in S^{p \times p}$ such that $(\alpha_i M_a, \gamma_i N_a)$ is a linear combination

$$(\alpha_i M_a, \gamma_i N_a) = \sum_{j=1}^{p} (R_a)_{ij} (\alpha_j, \gamma_j). \tag{1}$$

Also, $(\alpha, \gamma) \in \langle V_{k_0} \rangle$ yields that $(\alpha, \gamma)$ is a linear combination of the $(\alpha_j, \gamma_j)$:

$$(\alpha, \gamma) = \sum_{j=1}^{p} \kappa_j (\alpha_j, \gamma_j) \tag{2}$$

where $\kappa_j \in S$ for every $1 \leq j \leq p$. Now let

- $\kappa = (\kappa_1, \ldots, \kappa_p) \in S^{1 \times p}$ and
- $R = \sum_{a \in \Sigma} R_a a \in S\langle \Sigma \rangle^{p \times p}$.

Moreover, let $X$ be the $p \times m$ matrix over $S$ whose rows are the vectors $\alpha_1, \ldots, \alpha_p$ and $Y$ the $p \times n$ matrix over $S$ whose rows are the vectors $\gamma_1, \ldots, \gamma_p$. Moreover, let $\lambda = X\beta = Y\delta \in S^{p \times 1}$.

Then $\mathcal{C} = (\kappa, R, \lambda)$ is an automaton of dimension $p$. In addition, $X$ is a simulation $\mathcal{C} \to \mathcal{A}$ and $Y$ is a simulation $\mathcal{C} \to \mathcal{B}$. Indeed, $\kappa X = \alpha$ and $\kappa Y = \gamma$ by (2), and $R_a X = X M_a$ and $R_a Y = Y N_a$ for all $a \in \Sigma$ by (1). Finally, $X\beta = \lambda = Y\delta$. □

**Corollary 4.3.** *Suppose that every finitely generated subsemiring of a semiring $S$ is Noetherian. Then $S$ is proper.*

*Proof.* By Theorem 4.2 and Lemma 3.3. □

Since every finitely generated commutative ring is Noetherian, (cf. [2]), and since every finite semiring is clearly Noetherian, we obtain:

**Corollary 4.4.** *Every commutative ring is proper.*

**Corollary 4.5.** *Every finite semiring is proper.*

Call an automaton $(\alpha, M, \beta)$ in $S\langle\!\langle \Sigma^* \rangle\!\rangle$ (strictly) *deterministic* if $\alpha$ and each row of any $M_a$ for $a \in \Sigma$ is a *unit vector*; i.e., it has a single nonzero component which is 1. We end this section by pointing out that when $S$ is a finite semiring and $\mathcal{A} = (\alpha, M, \beta)$ and $\mathcal{B} = (\gamma, N, \delta)$ are equivalent automata in $S\langle\!\langle \Sigma^* \rangle\!\rangle$, then the automaton $\mathcal{C} = (\kappa, R, \lambda)$ in the proof of Theorem 4.2 can be chosen to be deterministic. Indeed, since $S$ is finite, there is some $k_0$ such that $V_{k_0}$ contains all vectors of the form $(\alpha_w, \gamma_w)$ with $w \in \Sigma^*$. We may choose $(\alpha_1, \gamma_1), \ldots, (\alpha_p, \gamma_p)$ to be an enumeration of $V_{k_0}$ with $(\alpha_1, \gamma_1) = (\alpha, \gamma)$, say. Now if $(\alpha_i M_a, \gamma_i N_a) = (\alpha_j, \gamma_j)$ for some $a \in \Sigma$ and $1 \leq i, j \leq p$, then we may define the $i$th row of $R_a$ to be the unit vector whose $j$th component is 1 (and whose other components are 0). Moreover, we can define $\kappa_1 = 1$ and $\kappa_i = 0$ for $1 < i \leq p$. The vector $\lambda$ is the same as before. We clearly have (1) and (2), so that the matrices $X$ and $Y$ are simulations $\mathcal{C} \to \mathcal{A}$ and $\mathcal{C} \to \mathcal{B}$, respectively.

In [4] a different argument is given to show that every finite positively ordered commutative semiring is proper. However, the argument applies to all finite semirings.

## 5 The tropical semiring

Below we will say that a semiring $S$ is *effectively presentable* if its carrier can be represented as a recursive subset of $\mathbb{N}$ such that its operations are recursive functions. Examples of effectively presentable semirings are all finite semirings, the semiring $\mathbb{N}$ and the *tropical semiring* [6, 7], which is $T = (\mathbb{N} \cup \{\infty\}, \min, +, \infty, 0)$ where the sum operation is the minimum, the product operation is ordinary addition with $n + \infty = \infty = \infty + n$ for all $n \in \mathbb{N} \cup \{\infty\}$, and the constants are $\infty$ and 0.

**Lemma 5.1.** *If $S$ is effectively presentable, then it is semidecidable whether two automata $\mathcal{A}$ and $\mathcal{B}$ in $S\langle\!\langle \Sigma^* \rangle\!\rangle$, where $\Sigma$ is a finite alphabet, are **not** equivalent.*

*Proof.* Let $w_0, w_1, \ldots$ be a recursive enumeration of $\Sigma^*$. For $i \geq 0$ compute $(|\mathcal{A}|, w_i)$ and $(|\mathcal{B}|, w_i)$. If $(|\mathcal{A}|, w_i) \neq (|\mathcal{B}|, w_i)$ for some $i$, then $\mathcal{A}$ and $\mathcal{B}$ are not equivalent. □

**Lemma 5.2.** *If $S$ is a proper and effectively presentable semiring, then it is semidecidable whether two finite automata in $S\langle\!\langle \Sigma^* \rangle\!\rangle$ over a finite alphabet $\Sigma$ are equivalent.*

*Proof.* Given two automata $\mathcal{A} = (\alpha, M, \beta)$ and $\mathcal{B} = (\gamma, N, \delta)$ in $S \langle\!\langle \Sigma^* \rangle\!\rangle$, we generate all finite sequences of automata $\mathcal{C}_0, \ldots, \mathcal{C}_k$ connecting $\mathcal{A}$ with $\mathcal{B}$ together with matrices $X_0, \ldots, X_{k-1}$ over $S$ of appropriate dimension. We check whether or not $X_i$ is a simulation $\mathcal{C}_{i-1} \to \mathcal{C}_i$ or $\mathcal{C}_i \to \mathcal{C}_{i-1}$ for each $i$. If a chain of simulations is found, then $\mathcal{A}$ and $\mathcal{B}$ are equivalent. □

From Lemmata 5.1 and 5.2 we can immediately conclude:

**Corollary 5.3.** *Suppose that $S$ is a proper and effectively presentable semiring. Then it is decidable whether two automata $\mathcal{A}$ and $\mathcal{B}$ in $S \langle\!\langle \Sigma^* \rangle\!\rangle$ are equivalent, where $\Sigma$ is any finite alphabet.*

**Theorem 5.4.** *The tropical semiring $T$ is not proper.*

*Proof.* Clearly, $T$ is an effectively presentable semiring. It was shown in [8] that the equivalence problem of automata in $T \langle\!\langle \Sigma^* \rangle\!\rangle$ is undecidable when $\Sigma$ is a finite alphabet with at least two letters. Thus, $T$ is not proper by Corollary 5.3. □

Other effectively presentable semirings that are not proper are the variants of the tropical semiring studied in [8]. When $S$ embeds in $S'$ and $S$ is not proper, then $S'$ is not proper either.

## 6 Conclusion

We have shown that when $S$ is a Noetherian semiring and $\Sigma$ is any finite alphabet, then two automata $\mathcal{A}$ and $\mathcal{B}$ in $S \langle\!\langle \Sigma^* \rangle\!\rangle$ are equivalent if and only if they can be connected by a finite chain of simulations. In fact, when $\mathcal{A}$ and $\mathcal{B}$ are equivalent, then an automaton $\mathcal{C}$ was found together with simulations $\mathcal{A} \leftarrow \mathcal{C} \to \mathcal{B}$. When $S$ is finite, then $\mathcal{C}$ may be chosen to be deterministic. Moreover, we have shown that when $S$ is the tropical semiring, then for any finite alphabet $\Sigma$ of size at least 2 there exist equivalent automata in $S \langle\!\langle \Sigma^* \rangle\!\rangle$ that cannot be connected by any finite chain of simulations.